\documentclass[12pt]{iopart}
\pdfoutput=1

\usepackage{graphicx}
\usepackage{hyperref}

\usepackage{iopams}

\usepackage{booktabs} 
\usepackage{multirow}
\usepackage{footnote} 

\makesavenoteenv{tabular}
\makesavenoteenv{table}
\everymath{\displaystyle}

\begin{document}

\title[Improving LIGO calibration accuracy from temporal variations]{Improving LIGO calibration accuracy by tracking and compensating for slow temporal variations}

\author{D~Tuyenbayev$^{1,2}$, S~Karki$^{1,3}$, J~Betzwieser$^{4}$, C~Cahillane$^{5}$, E~Goetz$^{1}$, K~Izumi$^{1}$, S~Kandhasamy$^{4,6}$, J~S~Kissel$^{1}$, G~Mendell$^{1}$, M~Wade$^{7}$, A~J~Weinstein$^{5}$, and R~L~Savage$^{1}$}

\medskip
\address{$^{1}$LIGO Hanford Observatory, Richland, WA 99352, USA }
\address{$^{2}$University of Texas Rio Grande Valley, Brownsville, TX 78520, USA }
\address{$^{3}$University of Oregon, Eugene, OR 97403, USA}
\address{$^{4}$LIGO Livingston Observatory, Livingston, LA 70754, USA}
\address{$^{5}$California Institute of Technology, Pasadena, CA 91125, USA}
\address{$^{6}$University of Mississippi, Oxford, MS 38677, USA}
\address{$^{7}$Kenyon College, Gambier, OH 43022, USA}

\eads{\mailto{darkhan.tuyenbayev@utrgv.edu}, \mailto{richard.savage@ligo.org}}

\date{Nov 20, 2016}

\begin{abstract}
Calibration of the second-generation LIGO interferometric gravitational-wave detectors employs a method that uses injected periodic modulations to track and compensate for slow temporal variations in the differential length response of the instruments. These detectors utilize feedback control loops to maintain resonance conditions by suppressing differential arm length variations.  We describe how the sensing and actuation functions of these servo loops are parameterized and how the slow variations in these parameters are quantified using the injected modulations.
We report the results of applying this method to the LIGO detectors and show that it significantly reduces systematic errors in their calibrated outputs.

\end{abstract}
%
\vspace{-5pt}
\pacs{04.80.Nn, 95.55.Ym, 42.62.-b}
{\vspace{5pt}
\begin{indented}
\item[]{\it Keywords\/}: calibration, advanced LIGO, time-dependent parameters, photon calibrator, gravitation-wave detector
\end{indented}}
%

\section{Introduction}
\label{sec:intro}

Gravitational wave (GW) detectors are instruments designed to detect and measure ripples in the geometry of spacetime caused by cataclysmic astrophysical events such as the inspiral and coalescence of binary neutron star or binary black hole systems \cite{PhysRevLett.116.061102, PhysRevLett.116.241103}. Ground-based gravitational wave detectors such as those of the Advanced Laser Interferometer Gravitational-wave Observatory (LIGO) project are km-scale dual-recycled Fabry--P\'{e}rot Michelson interferometers with relative displacement sensitivities of better than $10^{-19}$ $\mathrm{m/\sqrt{Hz}}$ for frequencies near 150 Hz \cite{detupgrade}.
Accurate calibration of the reconstructed gravitational wave strain signals projected onto the detectors is crucial for both the detection of GW signals and for the subsequent extraction of the parameters of the sources \cite{lindblom2009,P1500248}.

GWs cause apparent variations in the relative lengths of the interferometer arms. These variations are sensed as power fluctuations at the GW readout port of the interferometer. Feedback control loops actuate the interferometer mirror positions to maintain resonance in the optical cavities and the desired interference condition at the beamsplitter. Thus, the apparent arm length fluctuations caused by external disturbances such as gravitational waves are suppressed by the differential arm length (DARM) feedback control loop. Accurately reconstructing these \emph{external} arm length fluctuations from the error and control signals of the servo loop is one of the primary goals of the LIGO calibration effort.

For the Initial LIGO detectors, slow temporal variations were attributed to frequency-independent changes in the overall gain of the sensing function \cite{P0900120}. The Advanced LIGO interferometers are more sophisticated than earlier detectors \cite{detupgrade,P1400177}. Temporal variations of the sensing function of the Advanced LIGO detectors involve both a changing scalar gain factor and frequency dependent changes due to a varying coupled-cavity pole frequency \cite{mizuno1993,P1500277}. Additionally, the actuation function is time dependent due to slow variations in the strength of an electrostatic force actuator.

During the first observation period of Advanced LIGO, between September 2015 and January 2016 (O1), the DARM control loop time-dependent parameters were tracked at both LIGO detectors using the method described in this paper. Application of these parameters improve the agreement between measurements and models of actuation and sensing functions.
Applying corrections for the temporal variations improved the accuracy of the reconstructed differential arm length variations induced with photon radiation pressure from an auxiliary laser source.

This paper is organized as follows: systematic errors resulting from uncompensated variations in the sensing and actuation functions are discussed in section~\ref{sec:setup}. The method for tracking and compensating for temporal variations is described in section~\ref{sec:method}. The results of applying the method are presented in section~\ref{sec:results}. Conclusions are given in section~\ref{sec:conclusions}.

\section{Calibration errors due to slow temporal variations}
\label{sec:setup}

\begin{figure}[!ht]
\begin{indented}
\item[]%
\includegraphics[width = 0.8\linewidth]{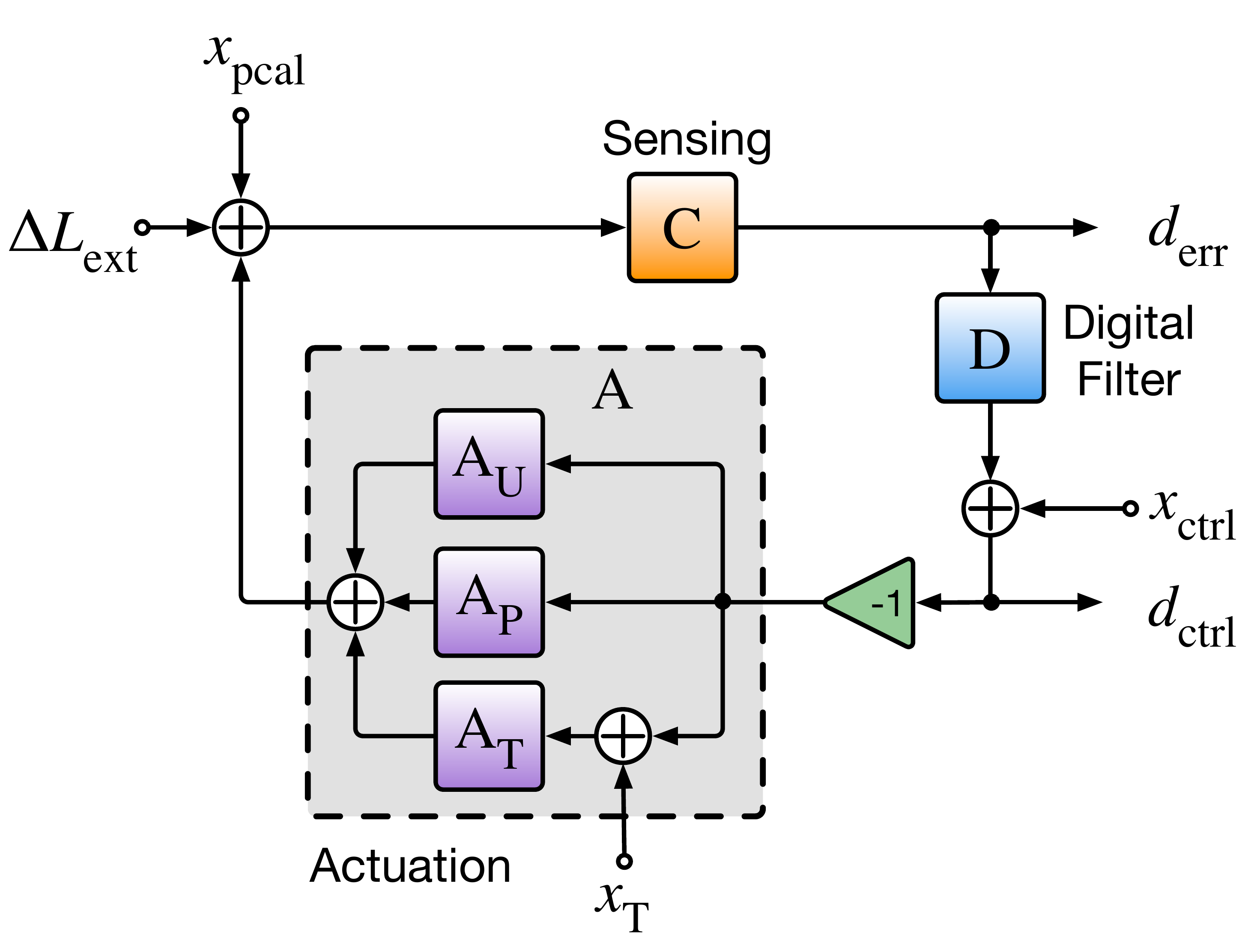}
\end{indented}
\caption{Schematic diagram of the differential arm length control loop. $C$ is the sensing function: response to changes in the apparent differential arm length; $D$ is the digital control filter transfer function; and $A_\mathrm{U}$, $A_\mathrm{P}$ and $A_\mathrm{T}$ are the actuation transfer functions of the upper-intermediate (U), penultimate (P) and test mass (T) stages of the quadruple pendulum suspension system. Differential arm length disturbances from sources outside (external) the control loop, e.g.\ GWs, are denoted by $\Delta L_\mathrm{ext}$. Injection points for modulated sinusoidal excitations (calibration lines) are denoted by: $x_\mathrm{pcal}$ -- excitations from a photon calibrator, $x_\mathrm{ctrl}$ -- excitations injected into the DARM control signal, and $x_\mathrm{T}$ -- excitations injected into the test mass actuation stage. $d_\mathrm{err}$ and $d_\mathrm{ctrl}$ represent the error and control signals of the loop.}
\label{fig:loopDiag}
\end{figure}

In the LIGO detectors, fluctuations in the differential arm length degree of freedom are suppressed by the DARM control loop. This servo is described in terms of a sensing function, $C(f, t)$, digital filters, $D(f)$, and an actuation function, $A(f, t)$, as shown in figure~\ref{fig:loopDiag}. A detailed discussion of the DARM loop is given in \cite{P1500248}. The response function of the detector, at any given time, $t$, is given by
\begin{equation}
	R(f, t) = \frac{1 + G(f, t)}{C(f, t)},
	\label{eq:R}
\end{equation}
where $G(f, t) = C(f, t) \, D(f) \, A(f, t)$ is the DARM open loop transfer function. The unity gain frequency is approximately $50\:\mathrm{Hz}$. Thus, the unsuppressed (external) differential arm length variations can be reconstructed from the DARM loop error signal by
\begin{equation}
	\Delta L_{\mathrm{ext},t'}(f) = \left. R(f, t) d_{\mathrm{err},t}(f) \: \right|_{t = t'}
	\label{eq:deltal_ext}
\end{equation}
The subscript ``$t$'' denotes that the quantity is a Fourier transform calculated over a short interval near time $t$.

Equations~(\ref{eq:R}) and (\ref{eq:deltal_ext}) show that systematic errors in the actuation and the sensing function models translate directly to systematic errors in the reconstructed $\Delta L_\mathrm{ext}$. Thus, it is important that the temporal variations in these functions are measured, and if possible compensated for, in calculation of $\Delta L_\mathrm{ext}$.

%
The sensing function of an Advanced LIGO interferometer includes the optical response of the signal recycled Fabry--P\'{e}rot Michelson interferometer and the frequency dependence of the output readout photodetector electronics \cite{P1500248}.
At time $t$ it is given by
\begin{equation}
	C(f, t) = \frac{\kappa_\mathrm{C}(t)}{1 + if / f_\mathrm{C}(t)} \, Q(f) \equiv S(f, t) \, Q(f),
	\label{eq:tdep_sens}
\end{equation}
where $Q(f)$ is the time-independent part of the sensing function that includes the photodetector response to laser power, responses of the electronics in the sensing chain, and the signal delay from the light travel time in the 4 km-long interferometer arms. $S(f, t)$ is the time-dependent part of the sensing function. It includes an optical gain scale factor, $\kappa_\mathrm{C}(t)$, and a coupled-cavity (the signal recycling and arm cavities) response of the interferometer, approximated by a single pole, $1 / \bigl( 1 + if / f_\mathrm{C}(t) \bigr)$.
The optical gain and coupled-cavity pole frequency vary due to slow drifts in the alignment and thermal state of the interferometer optics. While environmental effects, such as temperature fluctuation in the lab, cause alignment drifts, thermally distorted mirrors directly alter the spatial eigenmodes of the arm cavities and the signal recycling cavity resulting in mode-mismatch between these cavities. This, in turn, lowers the coupled-cavity pole frequency by reducing the signal recycling gain.

%
The test masses of an Advanced LIGO detector are suspended as the final stages of quadruple pendulum suspension systems \cite{quad2012}. The suspensions isolate the test masses from seismic disturbances and other environmental noise sources.
The DARM control loop uses the last three stages of the quadruple pendulum system: the upper-intermediate (U), penultimate (P) and the test mass (T) stages. The upper-intermediate and the penultimate stages use voice coil actuators, and the test mass stage uses an electrostatic force actuator (electrostatic driver, ESD).
The upper-intermediate stage actuators are dominant below $5\:\mathrm{Hz}$, the penultimate stage between $5$ and $20\:\mathrm{Hz}$ and the test mass stage above $20\:\mathrm{Hz}$. Details of actuation stage authority are discussed in greater detail in \cite{P1500248}.
The actuation function is the transfer function between a signal sent to the actuators and the induced displacement of the test mass at the end of a detector arm (end test mass, ETM).

%
%
The ESD actuation strength changes, apparently due to charge accumulation and due to drift in the bias voltage \cite{carbone2012, hewitson2007}.
The coil-magnet actuators used in the upper-intermediate mass and penultimate mass suspension stages, which are similar to actuators used in the Initial LIGO detectors \cite{P0900120}, are not expected to vary over time.
However, strengths of these actuators are tracked, regardless, in case of unexpected failures in their respective electronics chain.
Temporal variations in the actuation function model, $A(f,t)$, are parametrized with two scale factors: a test mass stage actuation scale factor, $\kappa_\mathrm{T}$, and a scale factor for the combined actuation functions of the penultimate and upper-intermediate stages, $\kappa_\mathrm{PU}$. Incorporating these scale factors, the actuation function is written as
\begin{equation}
	A(f, t) = \kappa_\mathrm{PU}(t) \bigl( A_\mathrm{P, 0}(f) + A_\mathrm{U, 0}(f) \bigr) + \kappa_\mathrm{T}(t) A_\mathrm{T, 0}(f),
	\label{eq:tdep_act}
\end{equation}
where $A_{\mathrm{P}, 0}(f)$, $A_{\mathrm{U}, 0}(f)$ and $A_{\mathrm{T}, 0}(f)$ are models of the actuation functions of the penultimate, upper-intermediate and the test mass stages. Here and throughout the paper the subscript ``0'' denotes that a function is evaluated at the reference time, $t_0$, when both $\kappa_\mathrm{PU}$ and $\kappa_\mathrm{T}$ are set to 1.


\begin{figure}[!ht]
\begin{indented}
\item[]%
\includegraphics[width = 0.8\textwidth]{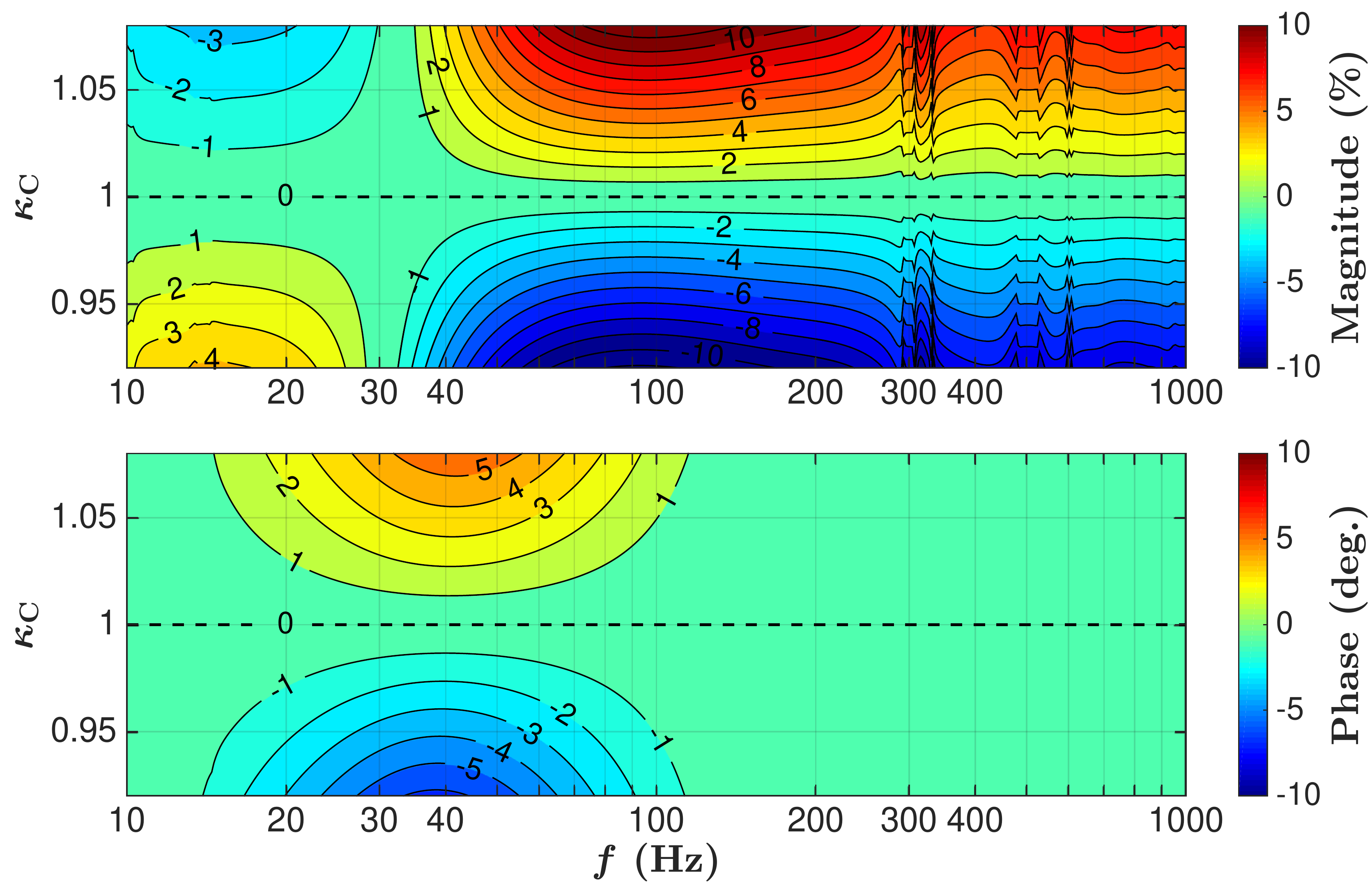}
\end{indented}
\caption{Estimated systematic calibration errors in the magnitude and phase of the response function resulting from uncorrected changes in the scale factor for the sensing function, $\kappa_\mathrm{C}$. Solid lines represent boundaries of $\pm 1~\%$, $\pm 2~\%$, $\pm 3~\%$, etc.}
\label{fig:kappa_c_carpet}
\end{figure}

\begin{figure}[!ht]
\begin{indented}
\item[]%
\includegraphics[width = 0.8\textwidth]{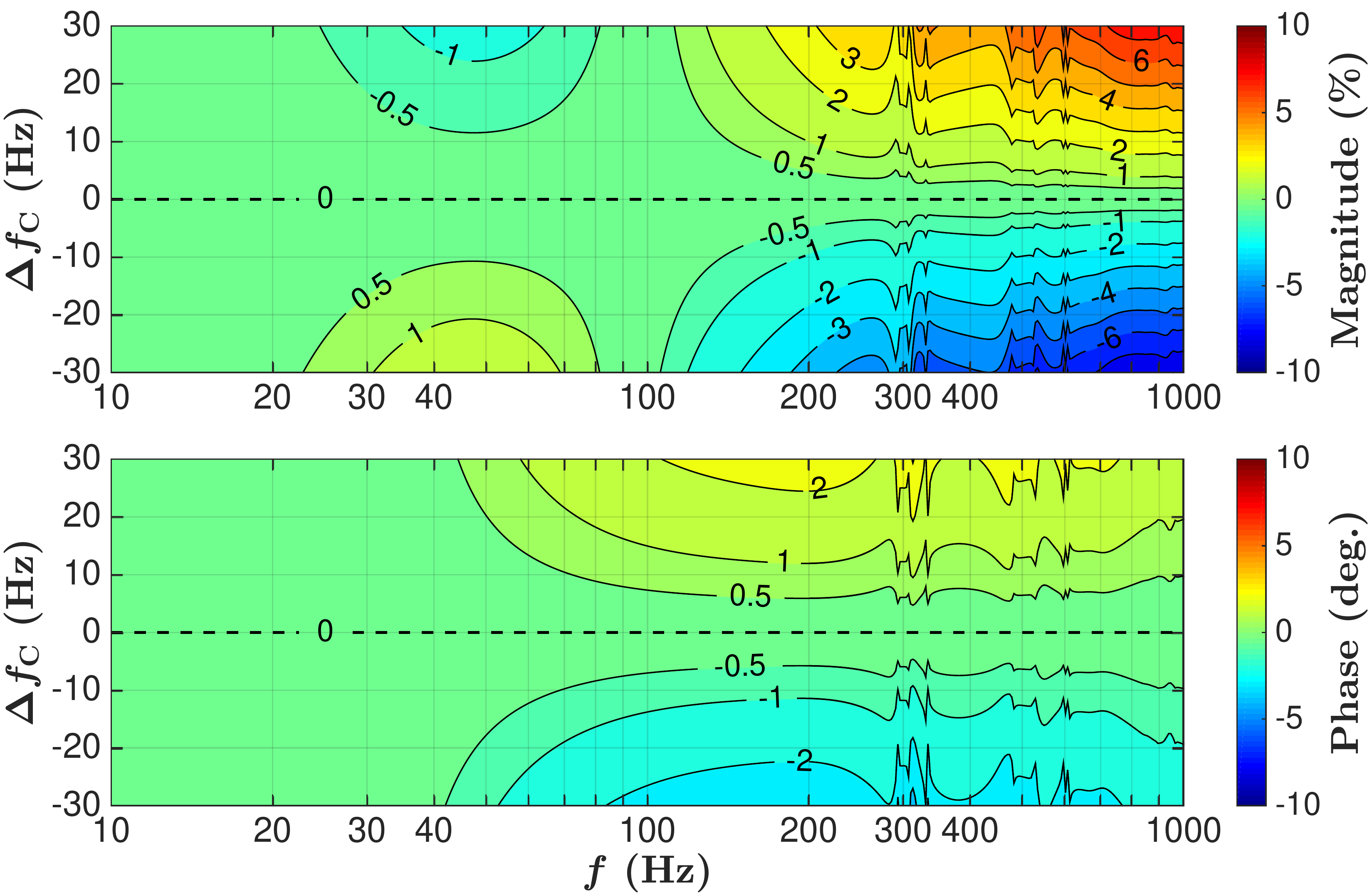}
\end{indented}
\caption{Estimated systematic calibration errors in the response function of the detector from uncorrected changes in the coupled cavity pole frequency, $\Delta f_\mathrm{C}$. Solid lines represent boundaries of $\pm 0.5~\%$, $\pm 1~\%$, $\pm 2~\%$, $\pm 3~\%$, etc.}
\label{fig:f_c_carpet}
\end{figure}

\begin{figure}[!ht]
\begin{indented}
\item[]%
\includegraphics[width = 0.8\textwidth]{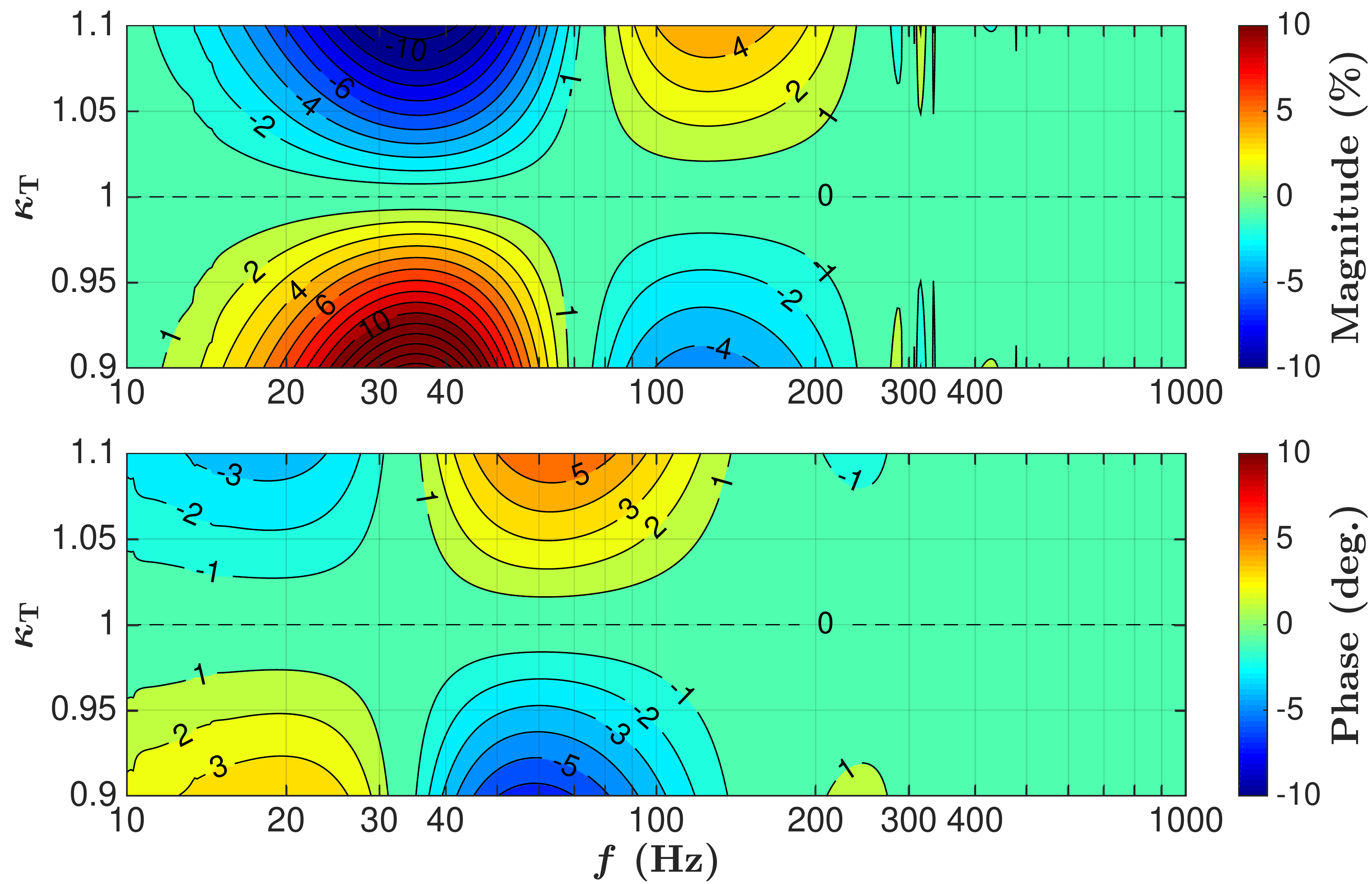}
\end{indented}
\caption{Estimated fractional systematic calibration errors from uncorrected scalar changes in the test mass stage actuation, $\kappa_\mathrm{T}$. Solid lines represent boundaries of $\pm 1~\%$, $\pm 2~\%$, $\pm 3~\%$, etc.\ systematic error regions and the case for the nominal value of $\kappa_\mathrm{T}$ is indicated with dashed line.}
\label{fig:kappa_tst_carpet}
\end{figure}

Not compensating for variations in the DARM control loop parameters can introduce systematic errors into the reconstruction of $\Delta L_\mathrm{ext}$.
These errors can be estimated by comparing a model of the response function of the detector in which loop parameters are varied to the same model with parameter values at the reference time.

Estimated systematic errors in the reconstruction of $\Delta L_\mathrm{ext}$ due to uncompensated changes in the sensing function scale factor and the coupled-cavity pole frequency (equation~(\ref{eq:tdep_sens})) are shown in figures~\ref{fig:kappa_c_carpet} and \ref{fig:f_c_carpet}, respectively.
The sensing function dominates $R(f)$ at higher frequencies where $|G| \ll 1$ (equation~(\ref{eq:R})), $R(f) \approx 1 / C(f)$. Therefore uncompensated changes in the sensing scale factor produce significant systematic errors at frequencies above the unity gain frequency (${\sim}50\:\mathrm{Hz}$). Changes in the coupled-cavity pole frequency produce significant systematic errors in the response function at frequencies near and above the coupled-cavity pole frequency (${\sim}340\:\mathrm{Hz}$).

At frequencies below the DARM loop unity gain frequency, where $|G| \gg 1$, $R(f) \approx A(f) \, D(f)$.
The actuation function, $A(f)$, is composed of three terms, one for each of the three suspension stages (see equation~(\ref{eq:tdep_act})).
Because $A_\mathrm{T}(f)$ is the dominant term at frequencies above $20\:\mathrm{Hz}$, systematic errors in $R(f)$ due to variations in the ESD actuation strength appear mostly in the frequency band from $20$ to $60\:\mathrm{Hz}$ as shown in figure~\ref{fig:kappa_tst_carpet}.

\section{Tracking and compensating for slow temporal variations}
\label{sec:method}

Temporal variations in the DARM control loop parameters can be monitored using modulated excitations injected into the DARM loop. These excitations produce peaks, or lines, at the modulation frequencies in the amplitude spectral density of the $d_\mathrm{err}$ signal.
The method for tracking temporal variations in the DARM control loop described in this paper requires monitoring the responses of the interferometer to four calibration lines injected into the DARM control loop: two lines injected using a photon calibrator system, $x_\mathrm{pcal}$, one line injected into the overall DARM actuation, $x_\mathrm{ctrl}$, and a line injected into the test mass stage actuation, $x_\mathrm{T}$.

The photon calibrator induces modulated displacements of the ETM via photon radiation pressure from a 1047 nm auxiliary laser source \cite{karki2016}. The induced displacements are suppressed by the DARM control loop (see figure~\ref{fig:loopDiag}). Thus, for any time $t'$, responses in $d_\mathrm{err}$ at photon calibrator line frequencies, $f_\mathrm{pcal1}$ and $f_\mathrm{pcal2}$, are given by
\begin{equation}
	d_{\mathrm{err},t'}(f_\mathrm{pcal1,2}) = \left.\frac{C(f, t)}{1+G(f, t)} x_{\mathrm{pcal},t}(f) \: \right|_{f = f_\mathrm{pcal1,2}, \, t = t'}
	\label{eq:calline_pcal}
\end{equation}


The lines injected into the overall DARM actuation control, $x_\mathrm{ctrl}$, and into the test mass stage actuation, $x_\mathrm{T}$, at frequencies $f_\mathrm{ctrl}$ and $f_\mathrm{T}$, will produce responses in the $d_\mathrm{err}$ signal that are also suppressed by the DARM control loop. These responses can be estimated as
\begin{eqnarray}
	d_{\mathrm{err},t'}(f_\mathrm{ctrl}) &= \left. \frac{-A(f, t) \, C(f, t)}{1+G(f, t)} x_{\mathrm{ctrl},t}(f) \: \right|_{f = f_\mathrm{ctrl}, \, t = t'} \label{eq:calline_ctrl} \\[1ex]
	d_{\mathrm{err},t'}(f_\mathrm{T}) &= \left. \frac{\kappa_\mathrm{T}(t) A_{\mathrm{T},0}(f) \, C(f, t)}{1+G(f, t)} x_{\mathrm{T},t}(f) \: \right|_{f = f_\mathrm{T}, \, t = t'} \label{eq:calline_tst}
\end{eqnarray}


Temporal variations in the test mass stage actuation scale factor, $\kappa_\mathrm{T}$, are tracked using the responses to the $x_\mathrm{pcal}$ and $x_\mathrm{T}$ lines in $d_\mathrm{err}$ at nearby frequencies. Taking the ratio of equation~(\ref{eq:calline_tst}) over equation~(\ref{eq:calline_pcal}) and solving for $\kappa_\mathrm{T}(t)$ gives
\begin{equation}
	\fl \kappa_\mathrm{T}(t) = \frac{1}{A_{\mathrm{T}, 0}(f_\mathrm{T})} \, \frac{d_{\mathrm{err},t}(f_\mathrm{T})}{x_{\mathrm{T},t}(f_\mathrm{T})} \left( \frac{d_{\mathrm{err},t}(f_\mathrm{pcal1})}{x_{\mathrm{pcal},t}(f_\mathrm{pcal1})} \right)^{-1} \frac{C_0(f_\mathrm{pcal1})}{1 + G_0(f_\mathrm{pcal1})} \left( \frac{C_0(f_\mathrm{T})}{1 + G_0(f_\mathrm{T})} \right)^{-1},
	\label{eq:kappa_T}
\end{equation}
where $C_0$ and $G_0$ are the sensing and DARM open loop transfer functions at the reference time $t = t_0$ and $x_\mathrm{pcal}$ is a calibrated length modulation induced by the photon calibrator.
The ratio between the DARM response function magnitudes at these two calibration line frequencies does not change appreciably (more than a fraction of a percent) for typical variations in DARM parameters. The last two terms in equation~(\ref{eq:kappa_T}) can therefore be evaluated at the reference time.


The stability of the upper-intermediate and penultimate actuation stages are monitored by tracking the combined scalar gain factor, $\kappa_\mathrm{PU}$,
\begin{equation}
	\kappa_\mathrm{PU}(t) = \frac{1}{A_{\mathrm{P}, 0}(f_\mathrm{ctrl}) + A_{\mathrm{U}, 0}(f_\mathrm{ctrl})} \Bigl( A(f_\mathrm{ctrl}, t) - \kappa_\mathrm{T}(t) A_{\mathrm{T}, 0}(f_\mathrm{ctrl}) \Bigr)
	\label{eq:kappa_pu}
\end{equation}
The overall actuation at frequency $f_\mathrm{ctrl}$ is calculated from the responses to $x_\mathrm{ctrl}$ line and the same $x_\mathrm{pcal}$ line that was used for estimation of $\kappa_\mathrm{T}(t)$:
\begin{equation}
	\fl A(f_\mathrm{ctrl}, t') = \left. - \frac{d_{\mathrm{err},t}(f_\mathrm{ctrl})}{x_{\mathrm{ctrl},t}(f_\mathrm{ctrl})} \left( \frac{d_{\mathrm{err},t}(f_\mathrm{pcal1})}{x_{\mathrm{pcal},t}(f_\mathrm{pcal1})} \right)^{-1} \frac{C_0(f_\mathrm{pcal1})}{1 + G_0(f_\mathrm{pcal1})} \left( \frac{C_0(f_\mathrm{ctrl})}{1 + G_0(f_\mathrm{ctrl})} \right)^{-1} \: \right|_{t = t'}
	\label{eq:A}
\end{equation}

%
To reduce systematic errors in the estimated $\kappa_\mathrm{T}$ the two calibration lines, $x_\mathrm{T}$ and $x_\mathrm{pcal}$, are placed at nearby frequencies. Similarly, reduction of systematic errors in $A(f_\mathrm{ctrl}, t)$, which is used in calculation of $\kappa_\mathrm{PU}$, requires placing the frequencies of the lines injected through $x_\mathrm{ctrl}$ and $x_\mathrm{pcal}$ close to each other. Thus all three calibration line frequencies for tracking temporal variations in the actuation function must be clustered in a narrow frequency band.
The frequency band near 35 Hz was chosen, because this is the frequency region where the magnitudes of the transfer functions of the combined penultimate and upper intermediate mass stage and the test mass stage are approximately equal, so that $\kappa_\mathrm{T}$ and $\kappa_\mathrm{PU}$ are calculated with similar uncertainties. Injecting the calibration lines at lower frequencies would require using a larger fraction of the available test mass stage actuation range because of the steep increase in the seismic noise \cite{PhysRevD.93.112004}.


The complex, time-dependent part of the sensing function can be calculated at the photon calibrator line frequency using its response function (equation~(\ref{eq:calline_pcal})) and the sensing function model (equation~(\ref{eq:tdep_sens})):
\begin{equation}
	S(f_\mathrm{pcal2}, t') = \left. \frac{1}{Q(f_\mathrm{pcal2})} \left( \frac{x_{\mathrm{pcal},t}(f_\mathrm{pcal2})}{d_{\mathrm{err},t}(f_\mathrm{pcal2})} - D(f_\mathrm{pcal2}) A(f_\mathrm{pcal2}, t) \right)^{-1} \: \right|_{t = t'}
	\label{eq:S}
\end{equation}
where $A(f_\mathrm{pcal2}, t)$ is the full DARM actuation function corrected with $\kappa_\mathrm{T}(t)$ and $\kappa_\mathrm{PU}(t)$.
Then $\kappa_\mathrm{C}(t)$ and $f_\mathrm{C}(t)$ can be written in terms of $S(f_\mathrm{pcal2}, t)$ as
\begin{eqnarray}
	\label{eq:kappaC}
    \kappa_\mathrm{C}(t) &= \: \frac{|S(f_\mathrm{pcal2}, t)|^2}{\mathfrak{R}[S(f_\mathrm{pcal2}, t)]}, \\[1ex]
	\label{eq:f_C}
    f_\mathrm{C}(t) &=- \frac{\mathfrak{R}[S(f_\mathrm{pcal2}, t)]}{\mathfrak{I}[S(f_\mathrm{pcal2}, t)]} f_\mathrm{pcal2}.
\end{eqnarray}

The choice of the photon calibrator line frequency for tracking temporal variations in the sensing function, $f_\mathrm{pcal2}$, is based on the strength of the response of $S(f, t)$ to variations in $\kappa_\mathrm{C}$ and $f_\mathrm{C}$, i.e.\ $ \partial S / \partial \kappa_\mathrm{C} $ and $ \partial S / \partial f_\mathrm{C} $ normalized to $|S(f, t)|$ at their respective frequencies. The definition of $S(f, t)$ (see equation~(\ref{eq:tdep_sens})) suggests that the precision of the estimated $\kappa_\mathrm{C}$ should not be affected by the choice of $f_\mathrm{pcal2}$, however the precision of the estimated $f_\mathrm{C}$ is maximized if $f_\mathrm{pcal2}$ is close to the nominal cavity pole frequency \cite{T1500533}.

Finally, the time-dependent parameter values and the time-domain models of the sensing and actuation functions can be used to reconstruct $\Delta L_\mathrm{ext}(t)$ from the DARM error signal as follows:
\begin{eqnarray}
	\Delta L_\mathrm{ext}(t) = &(\mathcal{P}_i(t) / \kappa_\mathrm{C}(t)) \ast \bigl( \mathcal{Q}_i \ast d_\mathrm{err}(t) \bigr) + \Bigl( \kappa_\mathrm{PU}(t) \bigl( \mathcal{A}_{\mathrm{P}, 0} + \mathcal{A}_{\mathrm{U}, 0} \bigr) \nonumber \\
	& + \kappa_\mathrm{T}(t) \mathcal{A}_{\mathrm{T}, 0} \Bigr) \ast \bigl( \mathcal{D} \ast d_\mathrm{err}(t) \bigr),
	\label{eq:deltal_ext_tdep}
\end{eqnarray}
where $\mathcal{P}_i(t)$ and $\mathcal{Q}_i$ are the time-domain filters created from inverses of the coupled cavity response, $1 + if / f_\mathrm{C}(t)$, and the time-independent part of the sensing function, $1/Q(f)$. $\mathcal{D}$, $\mathcal{A}_{\mathrm{P}, 0}$, $\mathcal{A}_{\mathrm{U}, 0}$ and $\mathcal{A}_{\mathrm{T}, 0}$ are time-domain filters created from a model of the digital filters and reference-time models of the actuation functions, and $\ast$ denotes convolution.

Note that $\mathcal{P}_i(t)$ is a function of time. Therefore, generating the $\Delta L_\mathrm{ext}(t)$ time-series, in which changes in all four time-dependent parameters are compensated, requires continuously updating the $\mathcal{P}_i(t)$ time-domain filter. Compensating for changes in scalar factors $\kappa_\mathrm{C}$, $\kappa_\mathrm{PU}$ and $\kappa_\mathrm{T}$ only can be accomplished using the $\mathcal{P}_i(t)$ filter created from the coupled-cavity response at the reference-time.

\section{Results}
\label{sec:results}

The method for tracking temporal variations in the DARM control loop described in this paper was implemented and evaluated using the Advanced LIGO detectors during their first observing run in the fall of 2015. In this section, we describe the performance of the method for tracking the DARM time-dependent parameters and applying the corrections.

As was discussed in section~\ref{sec:method}, the method requires injecting four calibration lines and monitoring their responses. Table~\ref{tbl:callines} lists the frequencies at which the lines were injected at the LIGO Hanford and LIGO Livingston detectors. The magnitudes of all four lines were set to give signal-to-noise ratios of 100 in 10-second Fourier transforms of the DARM error signal.

\begin{table}[!ht]
\caption{Calibration lines injected into the DARM control loop at the LIGO Hanford (H1) and LIGO Livingston (L1) detectors. Lines 1-3 are used for estimation of $\kappa_\mathrm{T}$ and $\kappa_\mathrm{PU}$, and line 4 for $\kappa_\mathrm{C}$ and $f_\mathrm{C}$.}
\vspace{0.5ex}
\begin{indented}
\item[]%
\begin{tabular*}{\linewidth}{ @{\extracolsep{\fill}} r c r r p{0.6\linewidth} }
\toprule
\multirow{2}{*}{\textbf{\#}} & \multirow{2}{*}{\textbf{Signal}} & \multicolumn{2}{c}{\textbf{Freq. (Hz)}} & \multicolumn{1}{c}{\multirow{2}{*}{\textbf{Line Purpose}}} \\ \cline{3-4}
 & & \multicolumn{1}{c}{\textbf{H1}} & \multicolumn{1}{c}{\textbf{L1}} \\
\midrule
1 & $x_\mathrm{T}$ & 35.9 & 35.3 & Test mass stage actuation strength, equation~(\ref{eq:kappa_T}).\\
2 & $x_\mathrm{pcal}$ & 36.7 & 34.7 & DARM actuation, equations~(\ref{eq:kappa_T}), (\ref{eq:kappa_pu}).\\
3 & $x_\mathrm{ctrl}$ & 37.3 & 33.7 & Strength of the combined penultimate and upper intermediate actuation, equation~(\ref{eq:kappa_pu}).\\
4 & $x_\mathrm{pcal}$ & 331.9 & 331.3 & Sensing scale factor and coupled-cavity pole frequency, equations~(\ref{eq:kappaC}), (\ref{eq:f_C}). \\
\bottomrule
\end{tabular*}
\label{tbl:callines}
\end{indented}
\end{table}

\begin{figure}[!ht]
\begin{indented}
\item[]%
\includegraphics[width=\linewidth]{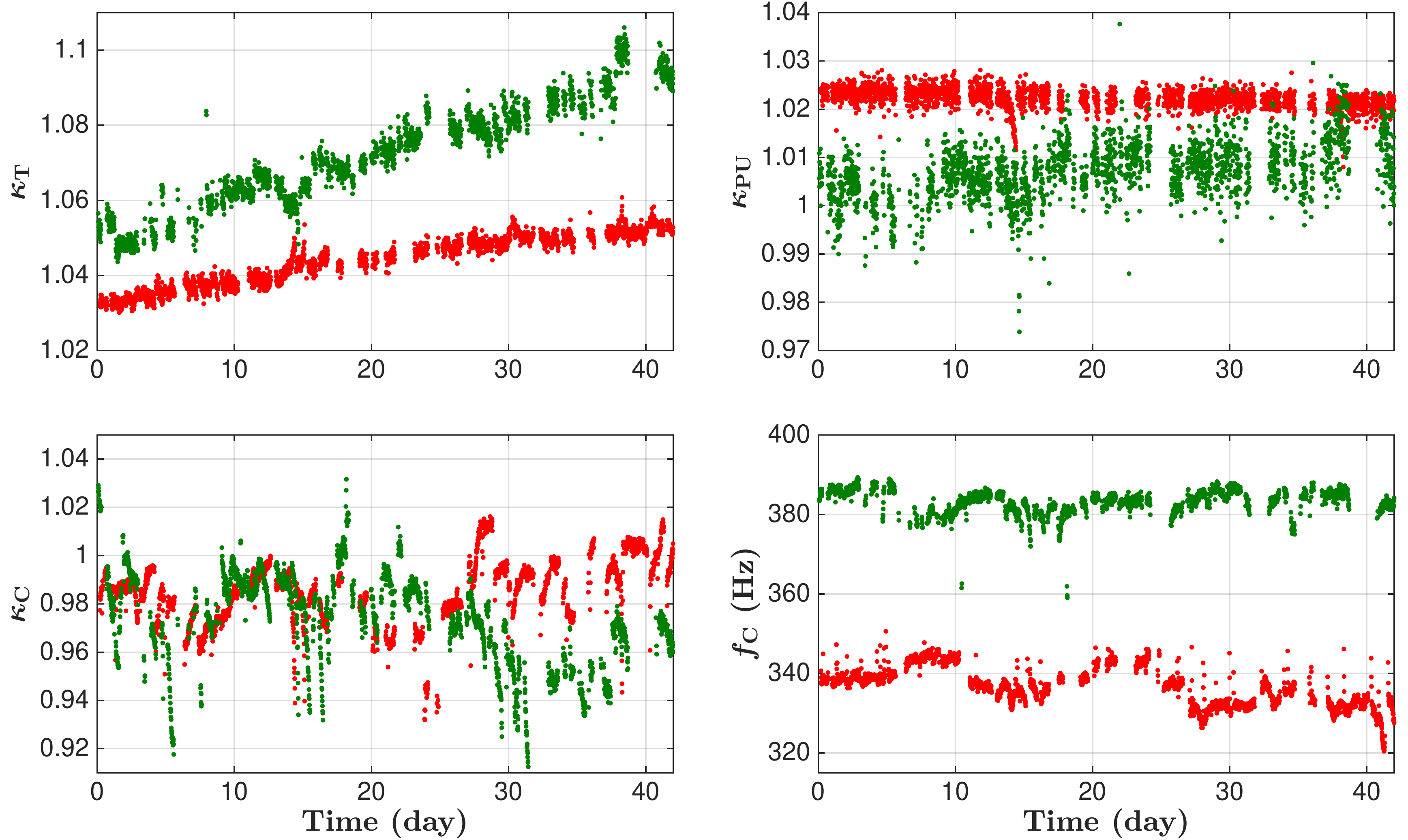}
\end{indented}
\caption{DARM time-dependent parameters calculated from calibration lines -- LIGO Hanford (\emph{red traces}) and LIGO Livingston (\emph{green traces}). Nominal values of all three scalar factors $\kappa_\mathrm{T}$, $\kappa_\mathrm{PU}$ and $\kappa_\mathrm{C}$ are 1, and the nominal value of the coupled cavity pole frequency, $f_\mathrm{C}$, for LIGO Hanford is $341\:\mathrm{Hz}$ and for LIGO Livingston is $388\:\mathrm{Hz}$ \cite{llo-alog-fc}.}
\label{fig:kappaTrends}
\vspace{10pt}
\end{figure}

\begin{figure}[!ht]
\begin{indented}
\item[]%
\includegraphics[width = \linewidth]{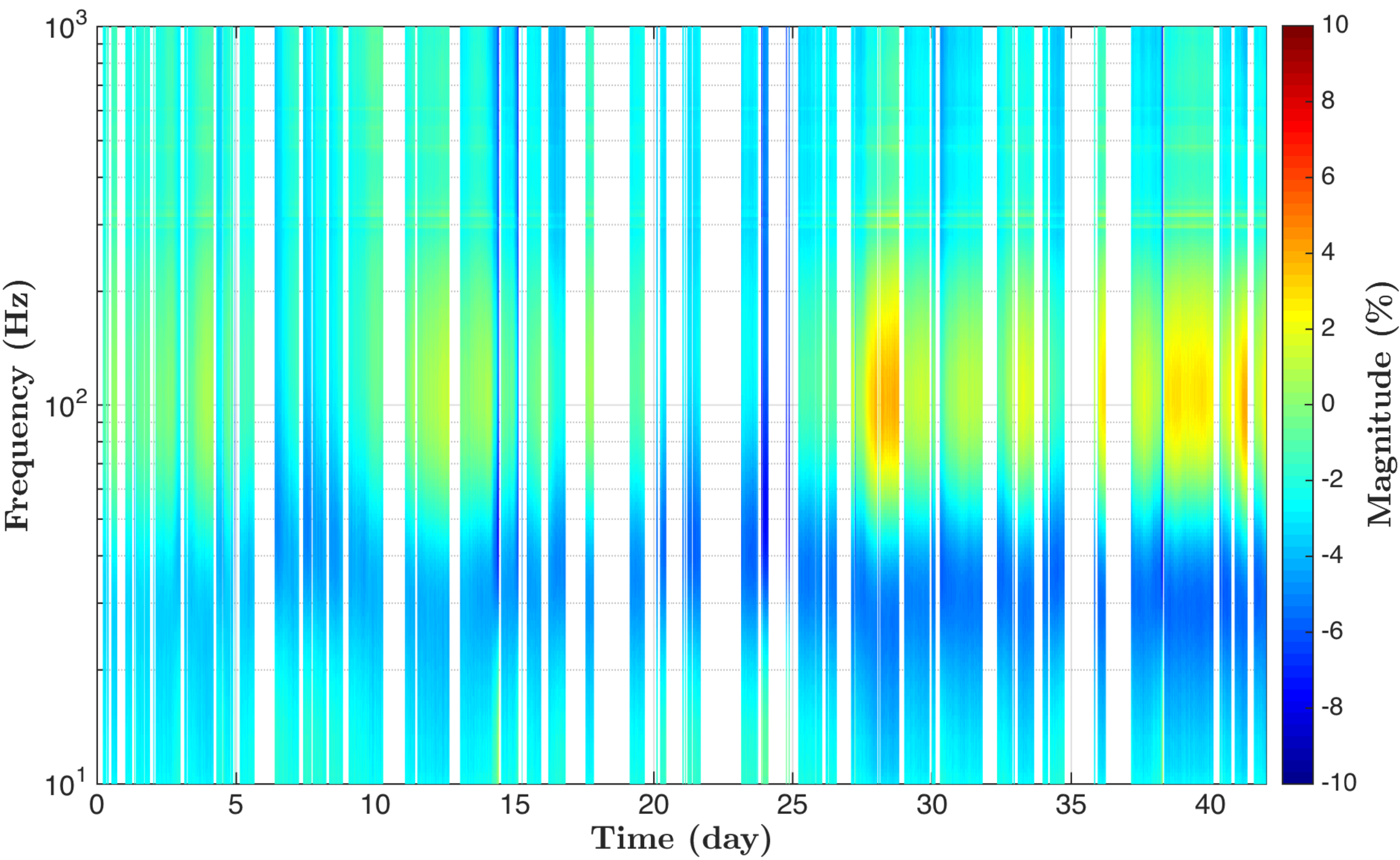}
\end{indented}
\caption{Time-dependent systematic errors in the static response function model of the LIGO Hanford detector, $R$, calculated using $\kappa_\mathrm{T}$, $\kappa_\mathrm{PU}$, $\kappa_\mathrm{C}$ and $f_\mathrm{C}$. The time spans 42 days in November and December 2015. The color axis represents systematic errors in percent.}
\label{fig:timeFreqDiagrams}
\vspace{10pt}
\end{figure}

\begin{figure}[!ht]
\begin{indented}
\item[]%
\includegraphics[width = \linewidth]{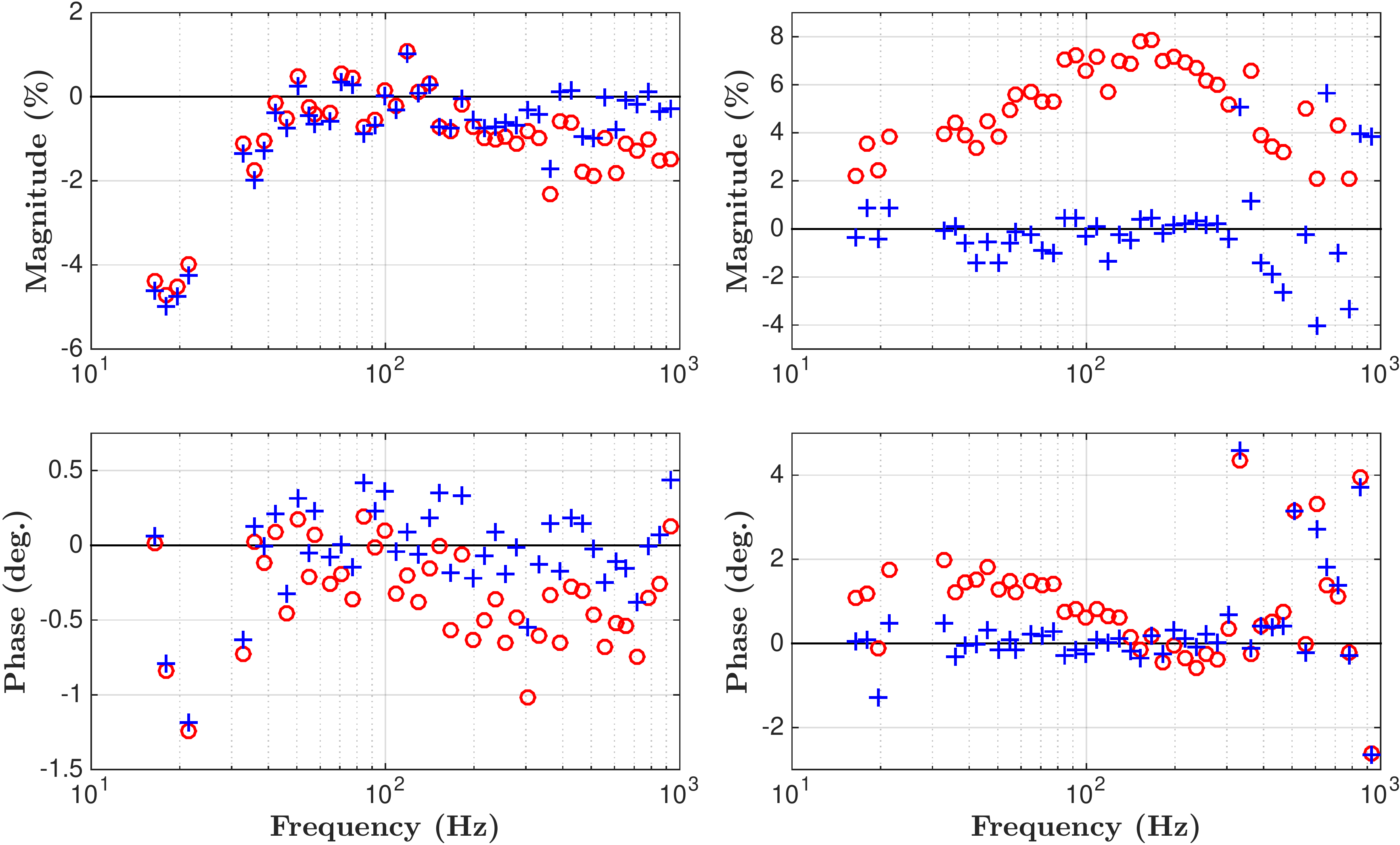}
\end{indented}
\caption{Deviation of measured sensing (\emph{left}) and actuation (\emph{right}) functions with respect to uncompensated reference-time models (\emph{red circles}) and models that incorporate time-dependent correction factors (\emph{blue plusses}).}
\label{fig:actAndSensTFs}
\vspace{10pt}
\end{figure}

\begin{figure}[!ht]
\begin{indented}
\item[]%
\includegraphics[width = \linewidth]{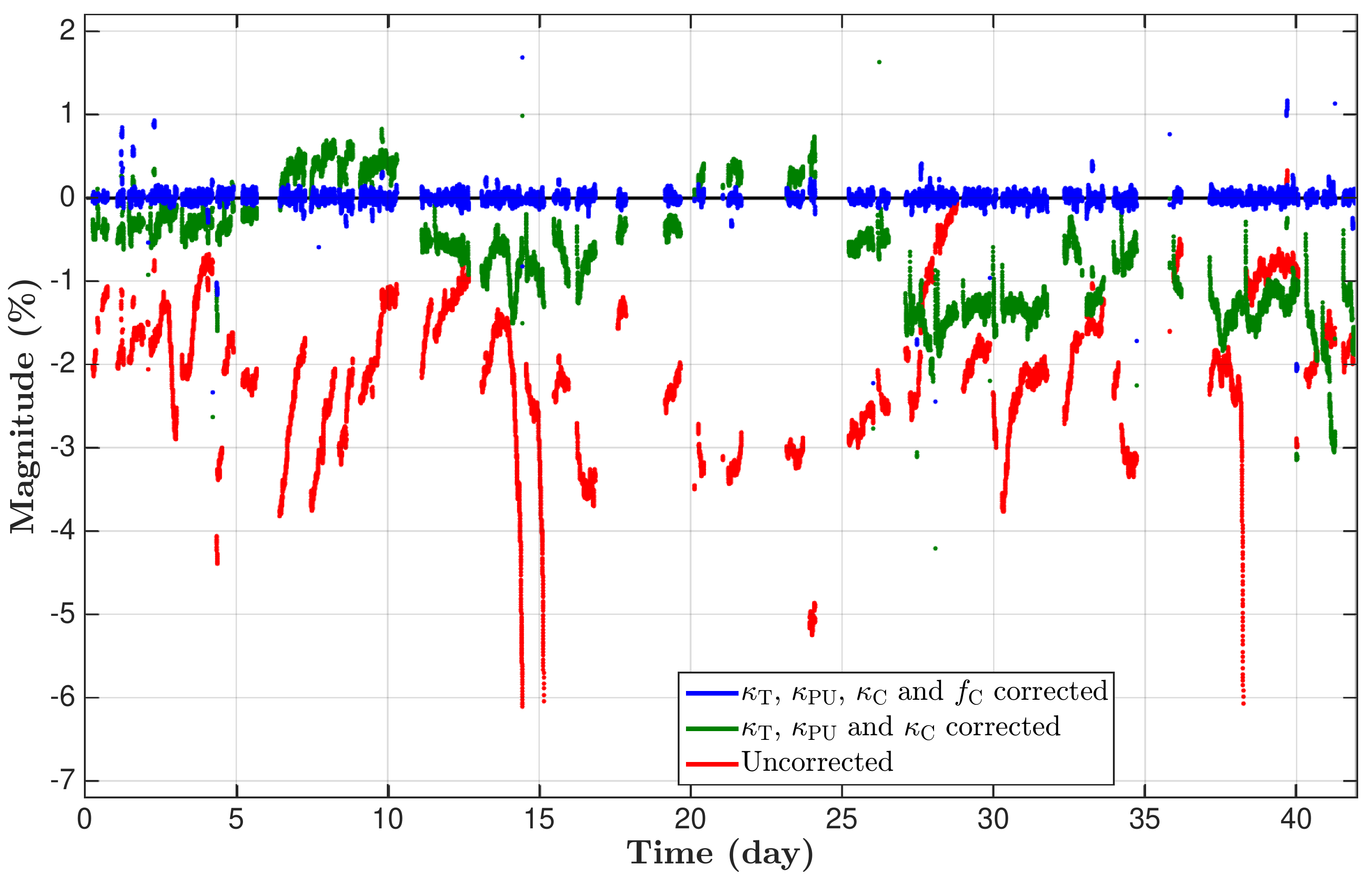}
\end{indented}
\caption{Systematic errors in the magnitude of $\Delta L_\mathrm{ext}$ reconstructed using static models of the sensing and actuation functions (\emph{red}), models with the parameters corrected for time-dependences in $\kappa_\mathrm{PU}(t)$, $\kappa_\mathrm{T}(t)$ and $\kappa_\mathrm{C}(t)$  (\emph{green}), and models that additionally include corrections for changes in the coupled-cavity pole frequency, $f_\mathrm{C}(t)$ (\emph{blue}). The data are averaged over 30 minute intervals.}
\label{fig:correctionSnippet}
\vspace{10pt}
\end{figure}

For both of the LIGO detectors the calculated values of the time-dependent parameters, $\kappa_\mathrm{T}(t)$, $\kappa_\mathrm{PU}(t)$, $\kappa_\mathrm{C}(t)$ and $f_\mathrm{C}(t)$, are shown in figure~\ref{fig:kappaTrends}.
These values can be used either to improve the estimation of external arm length fluctuations, as described in equation~(\ref{eq:deltal_ext_tdep}), or to evaluate time-dependent systematic errors in $\Delta L_\mathrm{ext}$ when the correction factors are not applied (see figure~\ref{fig:timeFreqDiagrams}).

The sensing and actuation function models are based on multiple-frequency sinusoidal excitation (swept-sine) measurements of the DARM open loop and the photon calibrator to $d_\mathrm{err}$ transfer functions at the reference time, $t_0$ \cite{P1500248, karki2016}. Frequency-dependent systematic errors in the models are estimated by comparing the subsequent swept-sine transfer function measurements of the sensing and actuation functions with reference-time models. Figure~\ref{fig:actAndSensTFs} shows how applying the time-dependent correction factors to the sensing and actuation models reduces the discrepancy between the measurements and the models. Correction factors were calculated from the calibration lines immediately before starting the transfer function measurements.

Tracking the high-frequency photon calibrator line amplitude in the reconstructed $\Delta L_\mathrm{ext}$ and comparing it to the displacement calculated from the photon calibrator readback signal indicates how slow temporal variations in the DARM control loop affect the calibration of the detector.
The photon calibrator line at $f_\mathrm{pcal2}$ was used to investigate the calibration accuracy of $\Delta L_\mathrm{ext}$ that was reconstructed using both the static sensing and actuation models and the models corrected with the time-dependent parameters. The results, averaged over 30 minutes, are shown in figure~\ref{fig:correctionSnippet}.
The data show that applying the scalar correction factors, $\kappa_\mathrm{T}$, $\kappa_\mathrm{PU}$ and $\kappa_\mathrm{C}$, significantly reduces the time-dependent systematic errors. During the first observation period of Advanced LIGO, reconstruction of the $\Delta L_\mathrm{ext}$ time-series incorporated corrections for variations in these scalar factors (\emph{green} data points).
Additionally applying corrections for the varying coupled-cavity pole frequency further reduces time-dependent systematic errors. As discussed at the end of section~\ref{sec:method}, correcting the $\Delta L_\mathrm{ext}$ time-series for variations in $f_\mathrm{C}$ requires continuously updating a time-domain filter. In the figure, the fully-corrected data (\emph{blue}) were generated by applying the coupled-cavity pole response calculated at a single frequency, $f_\mathrm{pcal2}$.
Figure~\ref{fig:correctionSnippet} shows that by using this method the systematic errors in the reconstructed $\Delta L_\mathrm{ext}$ can be reduced from as much as $6\,\%$ to below $1\,\%$.

\section{Conclusions}
\label{sec:conclusions}

The LIGO detectors rely on differential arm length (DARM) control loops to maintain desired resonances in optical cavities. The sensing and actuation functions of the control loops exhibit slow temporal variations. We have parametrized the temporal variations in the DARM loop with scalar factors for the test mass stage actuation, the combined penultimate and upper-intermediate stage actuation, an overall sensing scalar factor, and the coupled-cavity pole frequency of the sensing function. We have developed a method for tracking these temporal variations by monitoring the response of the DARM loop error signal to injected modulated displacements involving a photon calibrator, an electrostatic actuator and the overall DARM loop actuation.

Applying the time-dependent correction factors improves systematic errors in the magnitude of the reconstructed external differential arm length variations by several percent.

\section*{Acknowledgments}

LIGO was constructed by the California Institute of Technology and Massachusetts Institute of Technology with funding from the National Science Foundation (NSF) and operates under cooperative agreement PHY-0757058. This work was supported by the following NSF grants: HRD-1242090 for D.~Tuyenbayev, PHY-1607336 for S.~Karki, PHY-1404139 for S.~Kandhasamy and PHY-1607178 for M.~Wade.
Fellowship support for S.~Karki and D.~Tuyenbayev from the LIGO Laboratory and for D.~Tuyenbayev from the UTRGV College of Sciences are also gratefully acknowledged.
This paper carries LIGO Document Number LIGO-P1600063.

\section*{References}

\bibliographystyle{iopart-num-unurl}
\bibliography{P1600063_tdep}

\clearpage

\end{document}